\documentclass{cys}
\usepackage[utf8x]{inputenc}
\usepackage[spanish]{babel}
\usepackage{listings} 
\usepackage{mdwlist} 
\usepackage{graphicx}

\title{An Extension of Interactive Scores for Multimedia Scenarios
  with Temporal Relations for Micro and Macro Controls}

\author{Mauricio Toro$^1$, Myriam Desainte-Catherine$^2$, and Julien Castet$^2$}

\affil{ 
$^1$Universidad Eafit, Medellin, \authorcr
Colombia           
\authorcr \authorcr
$^2$Universit\'{e} de Bordeaux, LABRI, SCRIME, Bordeaux, \authorcr
France
\authorcr  \authorcr
}

\begin{document}

\maketitle

\begin{abstract}
Software to design multimedia scenarios is usually based either on a
fixed timeline or on cue lists,
but both models are unrelated temporally. On the contrary, the formalism of interactive scores
can describe multimedia scenarios with
flexible and fixed temporal relations among the objects of the
scenario, but cannot
express neither temporal relations for micro controls nor signal processing.
We extend interactive scores with such relations and with sound processing. We show some 
applications and we describe how they can be
implemented in Pure Data. Our implementation has 
low average relative jitter even under high \textsc{cpu} load.
\end{abstract}



\othertitle{Una extensi\'{o}n al formalismo de partituras interactivas para escenarios multimedia con relaciones temporales para micro y macro controles} 
 
 \begin{resumen} 
 Software para dise\~nar escenarios multimedia es, usualmente, basado en una l\'{i}nea de tiempo fija o en una lista de eventos, pero ambos modelos se encuentran sin relaciones temporales. Por el contrario, el formalismo de partituras musicales interactivas puede describir escenarios multimedia, con relaciones temporales de duraci\'{o}n fija y flexible, entre los objetos del escenario, pero tampoco puede expresar relaciones temporales para micro controles ni para para procesamiento de señales. En este art\'{i}culo presentamos una extensión con ese tipo de relaciones y con procesamiento de señales. Nosotros mostramos algunas aplicaciones y describimos c\'{o}mo pueden ser implementadas en Pure Data. Nuestras implementaciones tienen un jitter promedio bajo, aún con una alta carga de procesamiento en CPU.
 \end{resumen} 

\begin{palabrasclave} 
partituras musicales interactivas, multimedia, interacción, programación concurrente por restricciones, procesamiento de sonido.
\end{palabrasclave} 


\section{Introduction}

Multimedia scenarios --such as interactive theater performances, interactive
museum exhibitions and Electroacoustic music-- are usually designed and
controlled by computer programs. It is crucial that the
software to execute such scenarios preserve the \textit{macroform} and the \textit{microform}.
The macroform comprises the structure of the scenario (e.g., the
tempo and the duration of the
scenes, movements, parts and measures). The microform
comprises the operations with samples (e.g., micro delays,
articulation, and sound envelope). In this
paper we deal with the macroform of multimedia content, but only with the microform of sound.

\subsection{Problems}
There are four problems with most existing multimedia scenario software: (1) time models
are unrelated temporally, (2) they provide no hierarchy, (3) the
different time
scales are unrelated, and (4) schedulers are not appropriate for real-time.
In what follows we explain each of them.

The first problem is that software to design multimedia scenarios is usually based either on a
fixed timeline with a very precise script, such as \textit{Pro
  Tools}\footnote{http://www.avid.com/US/resources/digi-orientation}, or a more flexible script
using cue lists, such as the theater cue manager \textit{Qlab}\footnote{http://figure53.com/qlab/}.
Another software to design such scenarios is \textit{Ableton Live}\footnote{http://www.ableton.com/}. Live is often used in Electroacoustic
music and performing arts because it allows to use both the fixed timeline and the cue lists, but
the two time models are unrelated temporally.
In fact, most software provide only one time model or
they are unrelated temporally. 

The second problem is that most software do not provide a hierarchy to represent the temporal objects of the scenario. As an
 example, using a hierarchy, it
is possible to control the start or end of an object
by controlling those from its parent. In interactive
 music, Vickery argues that a hierarchy is useful to control
 higher-order parameters of the piece; for instance, to control the
 volume dynamics, instead of the volume of each note \cite{Vickery04}.


The third problem is that the different time scales are often unrelated and cannot be
controlled in the same tool. \textit{Discrete user gestures} (e.g.,
clicking the mouse), 
\textit{control events} (e.g., control messages) and \textit{sound processing} have different
sampling frequencies and computing models. As an example, the audio processing language
\textit{Csound}\footnote{http://www.csounds.com/} has three types of
variables with different sampling rates: instrument variables, control variables and audio
variables. 

As a consequence of having the time scales unrelated, it is difficult to 
associate, for instance, a human gesture to both control events and
signal processing parameters in Csound.
To control signal processing and control events by human
gestures, \textit{Max/\textsc{msp}} and \textit{Pure Data (Pd)}
\cite{max} are often used, but they do not provide an environment to
design scenarios.

The fourth problem is that the most \textit{soft real-time}
schedulers, for instance 
those from Pd and Max, control both
signals and control messages together and they do not support
parallelism, thus they often fail to deliver control messages at the
required time; for instance, when they work under high \textsc{cpu} load, which 
is common when they process video, 3D graphics and sound. We argue 
that in soft real-time, the usefulness of a result degrades after its
deadline, thereby degrading the system's quality of service; whereas
in \textit{hard real-time} missing a deadline is a total system
failure (e.g., flight control systems). We focus on soft real-time.

To solve the problem of scheduling and to write high-performance
\textit{digital signal processors (\textsc{dsp}s)} for
 Max and Pd, users often write C++ plugins to model  
loops and independent threads. C++ plugins solve part of the
problem, but the control
messages --for the input and output of these plugins-- are handled 
by Max or Pd's schedulers. 

Another solution for the scheduler problem --often used 
during
live performance-- is to open two or more instances of Max or Pd 
simultaneously, running different programs on each one. Nonetheless,
synchronization
 is usually done either manually during performance or by using \textit{Open Sound Control (\textsc{osc})}, which adds more complexity and latency.

\subsection{Practical and conceptual implications}
The description of a multimedia scenario requires a consistent
relationship between the representation of the scenario in the
composition environment and the execution. Artistic creation requires
a composition of events at different time scales. As an example, it is
easy to describe that a video begins when the second string of a 
guitar arpeggio starts, but how can we achieve it in practice if the
beginning of the notes of the arpeggio is controlled by the user?

The problem emerges at runtime. The example given above is very
simple, but under high \textsc{cpu} load, a system interruption at
the point of playing the arpeggio and the video can often lead to
desynchronization. Usually, these eventualities are not considered by
developers, as the quality of systems is evaluated according to an
average performance. Nonetheless, during performance, it is desired
that the system works well even under high \textsc{cpu} load. 

The
synchronization between the arpeggio and the video must be achieved
 in every execution. If it does not work for a performance, concert or show,
the system performance is not satisfactory. Usually, artists 
prefer that an event is canceled if the event is not going to be
properly synchronized with the other media. Users want a system that
ensures that the events are either launched as they were defined in the score or they are not produced.

It is difficult to ensure determinism in the execution of multimedia
processes (e.g., sound, video and 3D images). Some operating
system like \textit{\textsc{rt}
  Linux}\footnote{http://www.windriver.com/index.html} or
\textit{RedHawk}\footnote{http://real-time.ccur.com/concurrent\_redhawk\_linux.aspx}
include priority queues for processes to respect hard real-time
constraints; however, in common operating systems, the user does not
have this type of control. 

This paper proposes a system to declare temporal
constraints among multimedia processes that aims to ensure all
temporal relations between events in the macroform and the microform
of the scenario; however, our solution remains under the realm of soft
real-time.

\subsection{Interactive scores}
There is a formalism to link both
the fixed timeline and the cue list model. 
The formalism of \textit{interactive
  scores} 
was proposed at the beginning of the century to describe scenarios with
flexible and fixed temporal relations among temporal objects
\cite{aad07}. 
 Examples of temporal objects are sounds, videos and light
 controls.   The designer
can specify that a video is played strictly before a light show, as an
example of flexible temporal relations. The designer can also specify that a drum loop
starts three seconds after the video, or
between 10 and 15 seconds after, as an example
of fixed temporal relations. 

Interactive scores also include a hierarchy of temporal objects: An
object contained inside another object must start after the execution
of its father and must end before its father ends. 
In addition, by using fixed temporal relations on the first
level of the hierarchy, it is possible to express absolute execution
times for the events of the scenario.

A mathematical structural definition, an
abstract semantics, formal properties of the scenarios, and an
operational semantics 
of interactive scores was presented in \cite{tdcr12}.
 
The formalism of interactive scores has also encouraged the development of
 software. 
An implementation of interactive scores is \textit{Virage}, which has
been
used for performing arts \cite{virage}. Another one is \textit{Iscore}
\cite{aadc08}, used for composition of Electroacoustic music. 
Unfortunately, neither Virage nor iScore provide a satisfactory solution to control sound
processing in real-time.

Virage can control different devices by the means of the \textsc{osc} protocol
and it can be used to model, for instance, curves that change the
value of a \textsc{dsp} parameter for sound synthesis. Nonetheless, the values of these
curves are sent at the control-event frequency and, therefore, its users cannot
express temporal relations at the sound processing level;
for instance, that one sound starts 500 $\mu$s after another.

\subsection{Contributions}
In this paper, we propose an extension to the interactive scores
formalism to define \textsc{dsp}s for sound synthesis. 
This paper deals with the macrostructure of multimedia, but only
microstructure of sound, and does not consider the structure of
image, video or other media. 


To define the microform of sound, we define a new type of temporal relations meant for high precision;
for instance, to express micro delays. We also introduce dataflow
relations; 
for instance, how the audio recorded by a temporal object is
transferred
to another object to filter it, add a micro delay, and then,
send it to another temporal object
to be diffused. The designer may define two views of the scenario: one
for temporal relations and another one for dataflow (e.g.,
Fig. \ref{fig:MicroDelayScore}); otherwise, relations may overlap. 

We also propose an encoding of the scenario into
two
models that interact during performance. (1) A model based on the
\textit{Non-deterministic
Timed Concurrent Constraint} (\texttt{ntcc}) calculus \cite{ntcc} for 
concurrency, user interactions and temporal relations, and (2) a model
based upon \textit{Faust} \cite{faust} 
for sound processing and micro controls. The great advantage of having a
formal model is that we could prove properties (e.g., playability) and
predict the behavior of the system. In fact, the interoperability 
of  ntcc and Faust has already been sketched in previous works \cite{audiontcc, iclp2010}. 

The novelty of our approach is using
the constraints sent from \texttt{ntcc} to control Faust.
We tested our applications in Pd, although they could also 
be compiled for Max or as a standalone program 
since
both Faust and \texttt{ntcc} can be translated into C++ and Max. In
fact, the final goal of our research is to develop a standalone
program. 

In what follows we briefly describe \texttt{ntcc}, Faust, and how they interact together.

\subsection{Non-deterministic Timed Concurrent Constraint}

In the process calculus \texttt{ntcc}, a system is modeled 
in terms of processes adding to a common \textit{store} partial information on the value of variables. Concurrent processes synchronize by blocking until a piece of information can be deduced from the store.

\texttt{Ntcc} includes the notion of discrete time as a sequence of
time units. Each time unit  
starts with a (possibly empty) store supplied by the environment. Processes scheduled for that time unit are then run until quiescence. The resulting store is the output at that time unit. Residual processes might also result. These are scheduled for the next (or any future) time unit and computation starts all over again.
 \texttt{Ntcc} has been used in
many musical applications \cite{cc-chapter}. The 
semantics of \texttt{ntcc} processes are given in \cite{ntcc}.

Process calculi has been applied to the modeling of interactive music systems
 \cite{is-chapter,tdcr14,ntccrt,cc-chapter,torophd,torobsc,Toro-Bermudez10,Toro15,ArandaAOPRTV09,tdcc12,toro-report09,tdc10,tdcb10,tororeport} 
 and ecological systems \cite{PT13,TPSK14,PTA13,mean-field-techreport}. 

A model for interactive scores based upon \texttt{ntcc} is proposed in
\cite{tdcr12}. In this model, the store contains all
the constraints from the temporal relations and the information of the
events launched by the user. Temporal
object processes synchronize themselves with the store. 

\texttt{Ntcc} models can be simulated in a real-time setting using
\textit{Ntccrt} \cite{ntccrt}. Ntccrt is based on \textit{Gecode} \cite{gecode}: state-of-the-art in constraint propagation.
Ntccrt programs can be compiled into standalone programs, or 
 plugins for Pd or Max.
Users can use Pd to
communicate any object with the Ntccrt plugin. In fact, Ntccrt can control
all the available objects for audio processing defined in
Pd, although our goal is to use Faust for
such tasks.

\subsection{Functional Audio STream (Faust)}

\begin{figure}[!h]
 \centerline{
 \includegraphics[width=\columnwidth]{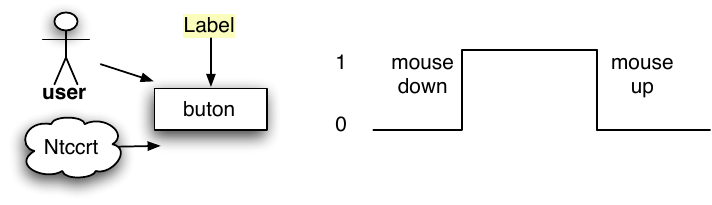}}
 \caption{The signal delivered by the button reflects the user actions
   (or the Ntccrt output): one when the button is pressed; zero otherwise.}
 \label{fig:FaustButon}
\end{figure}

Faust is a functional programming language for sound processing. In
Faust, \textsc{dsp} algorithms are functions
operating on signals. Faust programs are compiled into efficient C++
code 
that can be used in multiple programming languages and environments
\cite{AG07}. \textit{Graphical user interface (\textsc{gui})} objects in Faust can be defined in the same way as other signals.  We can control buttons, check boxes and integer inputs --originally designed for users--
from Ntccrt (Fig. \ref{fig:FaustButon}).


Although Faust programs can be compiled into efficient C++ programs,
Faust programs are limited because all signals must have the same sampling
rate. For that reason, 
Faust was recently extended for multirate \cite{jo11}. With such an 
extension, Faust would be capable to handle signals at different
frequencies. 
This is useful, for instance, for scenarios with different media such
as audio and video. Unfortunately, this extension is not yet
implemented, thus we only focus on sound processing.

Another extension of Faust is the \textit{Pd-Faust} interface \cite{AG07}. This interface
is useful for \textsc{dsp}s that cannot be efficiently implemented in Pd
because of a restriction of Pd: the 1-block minimum delay for feedback
loops. An example of such a \textsc{dsp} is the \textit{Karplus-Strong}
algorithm \cite{faust}. Furthermore, 
Pd-Faust can also be used for other \textsc{dsp}s. 

Finally, there is another reason to choose Faust: its extension 
for automatic parallelization and vectorization 
\cite{lof10}. This extension has been proved to be very efficient; for instance,
for the Karplus-Strong which we will use in several examples in
this paper. Orlarey \textit{et al.} found that using automatic
parallelization, a program that simulates simultaneously 32 strings
based on Karplus-Strong is twice faster using automatic
parallelization \cite{lof10}.


\subsection{Faust and ntcc interoperability}

Ntcc can send constraints to Faust, but currently Faust cannot send
information to ntcc because it requires subsampling. 
The constraints sent from ntcc cannot be partial information,
such as $pitch > 3$ or $gain < 1$ because such information cannot
be processed by Faust. Constraints must be equalities of the form
$variable = constant$. Using Pure Data, we can communicate those
values from ntcc to Faust by the means of number fields. 

\begin{figure}[!h]
 \centerline{
 \includegraphics[width=\columnwidth]{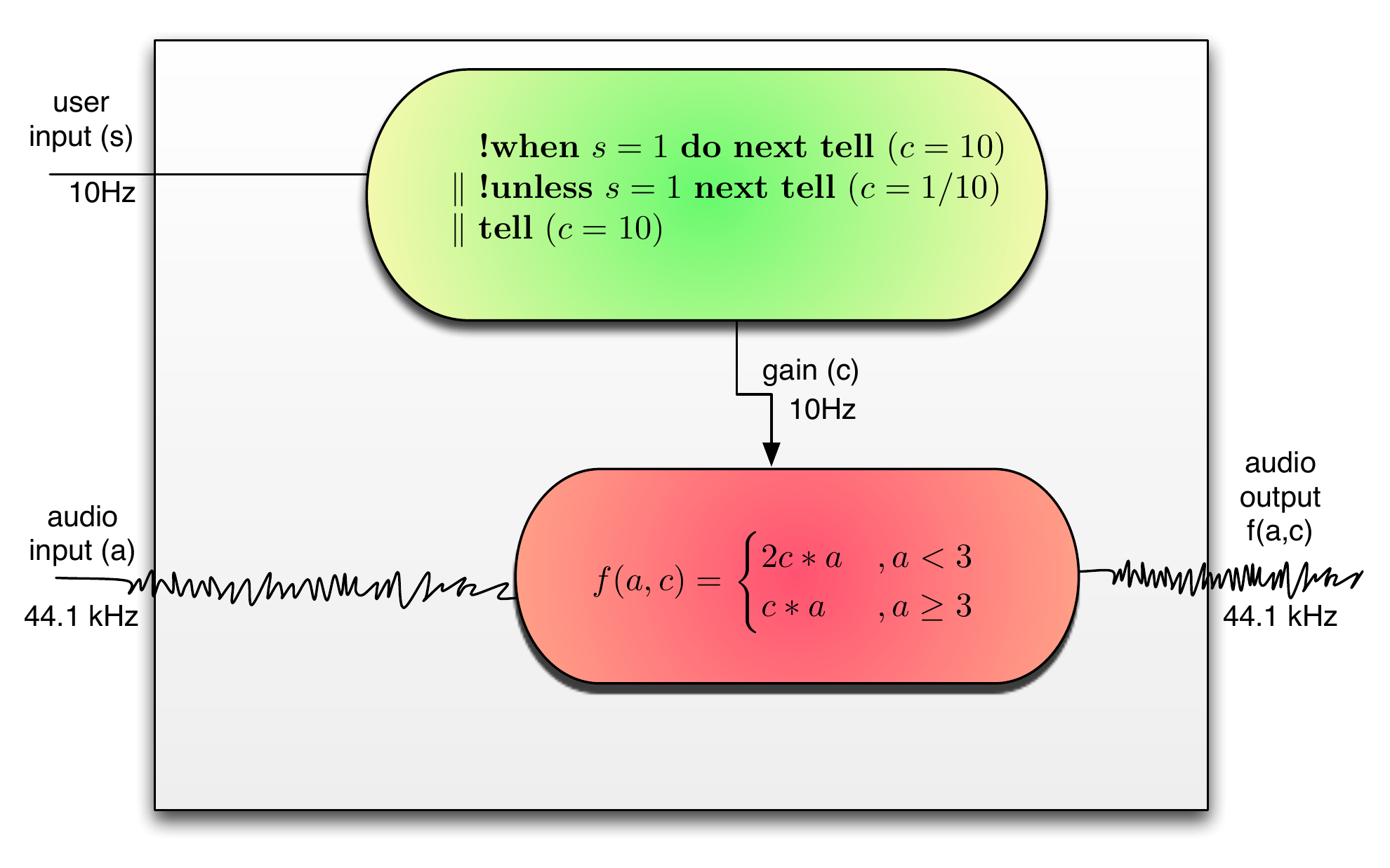}}
 \caption{Example of \texttt{ntcc} and Faust interoperability.}
 \label{fig:heterogeneousExample}
\end{figure}

As an example, we present a possible interoperoperation between Faust and ntcc
in Figure \ref{fig:heterogeneousExample}. On the one hand, ntcc can receive a user
input 
each discrete time unit. If the value of the input is $1$, ntcc
communicates to Faust that the gain is $10$; otherwise, if the user
gives no input, ntcc communicates Faust that the gain is $1/10$. On
the other hand, Faust
takes an audio signal an multiplies by the gain value given by
ntcc. In addition, Faust multiplies the signal by $2$ if the current
value of the audio input is
less than $3$. 

Note that ntcc cannot take decisions based on the values of the audio
signal 
because ntcc is not mean to handle audio signals, and Faust cannot
take decision based on absence of information or partial information.

\subsection{Structure of the Paper}
It is out of the scope of this paper to define formal semantics of
interactive scores. Semantics of interactive scores were defined in
\cite{tdcr12}. 
 It also out of the scope of this paper to fully describe the
semantics of interactive scores and Faust interoperability. Such a semantics is to be
defined in the
\textit{interactivity in the
writing of time and interactions} (\textsc{inedit}) project supported by
the \textit{french research agency} (\textsc{anr}). The purpose of
this 
project is 
to explore the interoperability of interactive scores, Faust, and other 
 french computer music software. The project will start
 in fall 2012. This paper offers preliminary and encouraging 
 results for \textsc{inedit}.

In what follows, we present the extension of interactive scores in
Section \ref{extension}; some applications developed with our framework in Section
\ref{applications}; quantitative results of the execution of the
application in Section \ref{results}; 
and conclusions, results and future work in Section \ref{conclusions}.

\section{Interactive Scores with Micro and Macro Controls}
\label{extension}
Scenarios in interactive scores are represented by \textit{temporal objects},
\textit{temporal relations for micro and
  macro controls}, \textit{interactive objects} and \textit{dataflow relations}.

\subsection{Temporal Objects}

Temporal objects can be triggered by interactive objects
 (usually launched by the user) and several temporal objects can be
 active simultaneously. The duration of a temporal object is given by
 an interval of natural numbers (which may include $\infty$). A temporal object may contain other temporal
 objects: this hierarchy allows us to control the start or end of a
 temporal object by controlling the start or end of its parent. 

Objects that do not have children, may have a sound synthesis process. A process is a
Faust program that is active during the execution of the object. 
These processes include at least two input signals: to control its
start and end. During the execution of a score, only one instance of a
temporal object can be active simultaneously because scores are linear
and loops are not considered in this extension. 

\subsection{Temporal Relations}
Temporal relations provide a partial order for the execution of the temporal objects; for instance, to express precedence between two objects.
In interactive scores, it is also possible to specify a variety of relations
among temporal objects such as 
global constraints and conditional branching. 

In this paper, we take into account scenarios limited to hierarchical
relations represented as a directed tree, \textit{point-to-point temporal
relations} without disjunction nor inequality ($\neq$),  
and quantitative temporal relations \cite{gennary98}.  The first 
\texttt{ntcc} model proposed in \cite{AADR06} is based on
Allen's relations; fortunately, point-to-point relations
can express all Allen's relations without disjunction
\cite{allen83}. We proposed a \texttt{ntcc} model with point-to-point
relations in \cite{tdcr12}. In this paper, we extend such a model to control Faust from
\texttt{ntcc}. 

In the model in \cite{tdcr12}, the relations between the start or end of two temporal
objects are labeled with an interval of integers that represents the possible
duration between the two points. Using $\infty$ in such intervals, it is possible to represent
the relations $<,>,\leq,\geq$ and $=$ with their usual interpretation over natural numbers.


In this paper, we also include \textit{high-precision temporal
  relations}. 
This new type of temporal relations between sound objects are meant to
have higher precision and they are controlled by Faust.
Temporal relations for sound-processing micro controls are labeled by an integer
$n$, where $n$ represents, for instance, a number of samples or
microseconds. Nonetheless, we can also use this relations to represent
durations of seconds. 
We represent graphically such relations with dashed arrows.

\subsection{Dataflow Relations}
A dataflow relation between objects $a$ and $b$ means that the audio outputs of $a$
are connected to the audio inputs of $b$. If $a$ has more outputs than $b$
inputs, they are merged; if $a$ has less outputs than $b$, they are
split. The control inputs of a Faust subprocess are connected
automatically depending on the dataflow, and the micro and macro controls.

\begin{figure}[!h]
 \centerline{
 \includegraphics[width=\columnwidth]{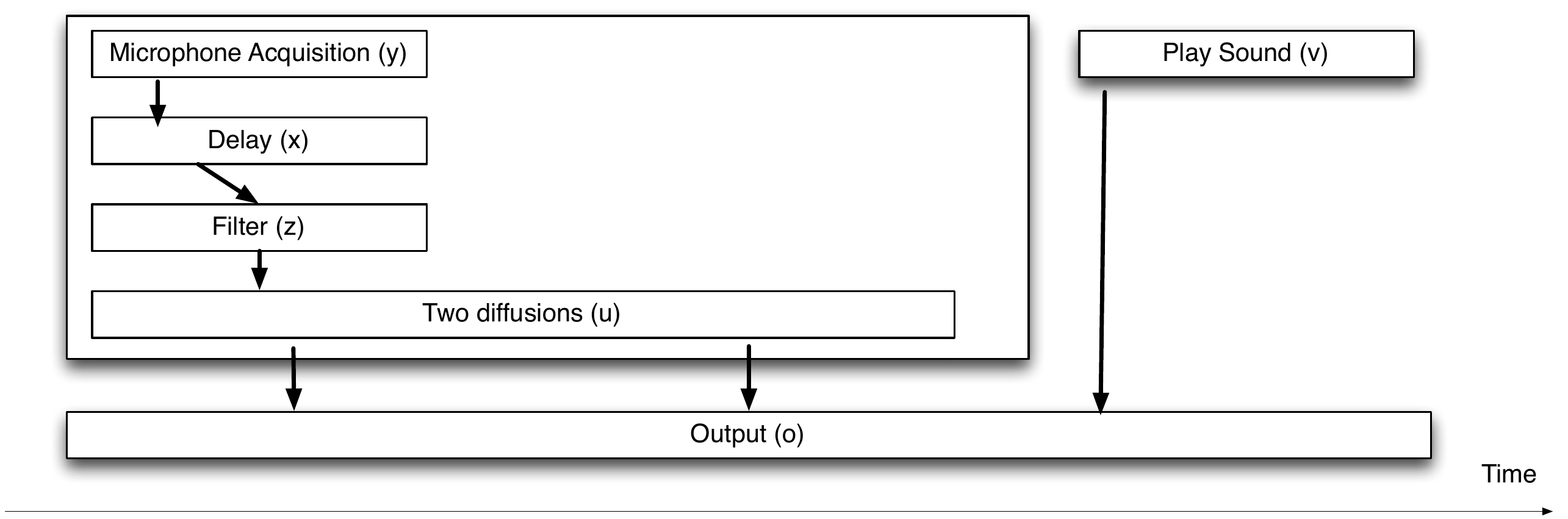}}
 \caption{\textit{Dataflow view} of a scenario. Thick arrows represent the flow of data through time.}
 \label{fig:dataflowExample}
\end{figure}

As an example, the reader may see the \textit{dataflow view} of a
scenario in Figure \ref{fig:dataflowExample}. In such a scenario, a sound is recorded by the
acquisition object, then the stream is passed to a delay object, and
then is passed to
a filter that adds gain. Finally, the stream is passed to an 
object that sends two copies of the stream to the output. In what
follow, we describe another example. 

\subsection{Example: An Arpeggio with Three Strings}
Karplus-Strong is an algorithm to generate metallic plucked-string
sounds. It can be described in a few lines of Faust. In
the
Faust program presented by Orlarey \textit{et al.} in \cite{faust}, a button 
triggers the sound. We connect  
such button to a control signal sent from the Ntccrt plugin to the Faust
plugin at the beginning of the temporal object. We also add another button to
stop the sound of the string. In Pure Data (Pd), such buttons
can be represented by
\textit{bang} or \textit{toggle} objects that send messages to the
plugin. In addition, we can use number fields as input for Faust. 
 We use Pd for simplicity, but Pd is not required to integrate Ntccrt with Faust.

\begin{figure}[!h]
 \centerline{
 \includegraphics[width=\columnwidth]{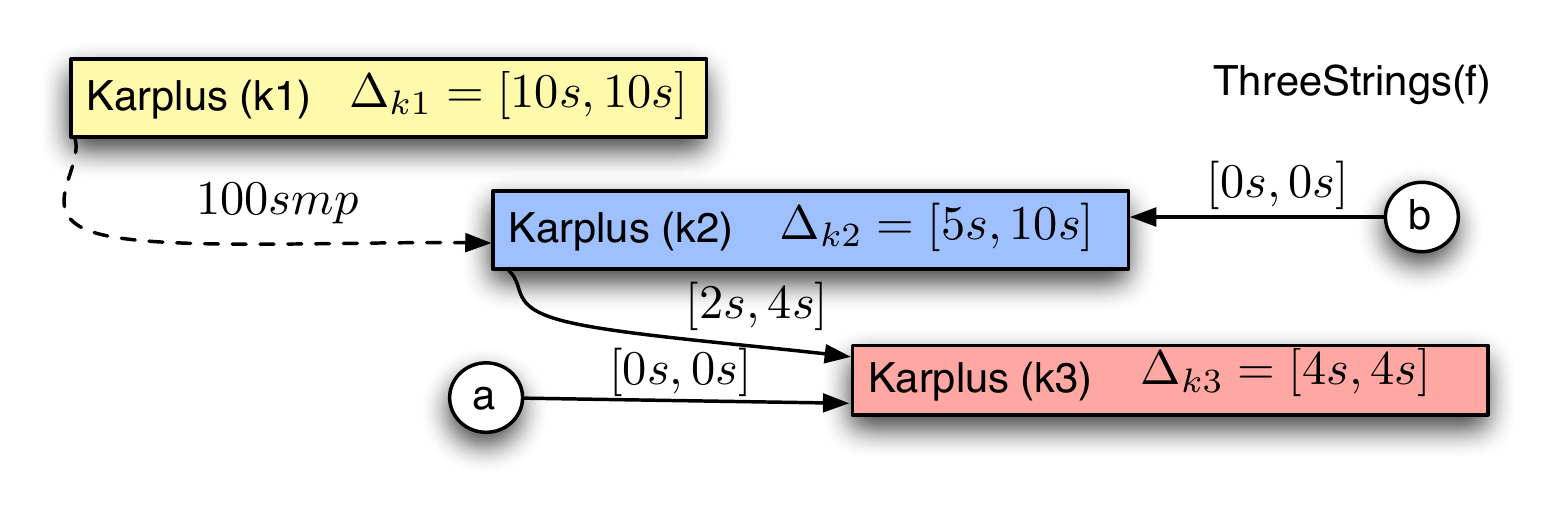}}
 \caption{An example of a scenario. The durations in the temporal relations are labeled
   with seconds ($s$) and in the high precision temporal relations with samples
   ($smp$). Interactive objects are $a$ and $b$.}
 \label{fig:threeFaust}
\end{figure}

Figure \ref{fig:threeFaust} is a scenario that models an arpeggio of
three strings using Karplus-Strong. The
dataflow is simple: each audio outputs is merged into a single 
output. There are two types of temporal relations: some labeled with  
intervals in the order of seconds that will be handled by the
\textit{Ntccrt} plugin, and the high precision ones, in the order of samples, that will be
handled by the Faust plugin.

The temporal constraints of the scenario are obtained from the duration
of each temporal object, the hierarchy and from the temporal
relations. For each temporal object, we add to the constraints: (1) ``the start time
of the object plus its duration is equal to the end time of the
object'' and (2) ``the object starts after its father and ends before its father''. For each temporal relation, we add the constraint ``the time
of the first point plus the duration in the relation is the time of
the second point''. The temporal constraints of the score are
explained in detail in \cite{tdcr12}. 

\begin{figure}[!h]
 \centerline{
 \includegraphics[width=\columnwidth]{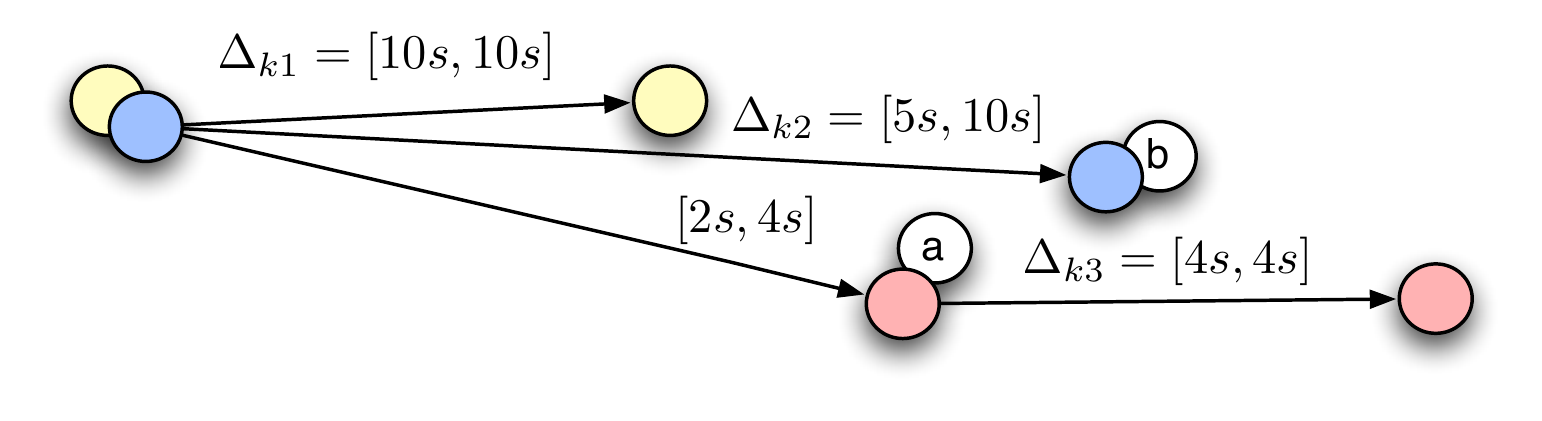}}
 \caption{The temporal constraints of the scenario in Figure
   \ref{fig:threeFaust}. The durations in the temporal relations are labeled
   with seconds ($s$) and the high precision temporal relations are
   considered as zero delays.}
 \label{fig:ConstraintGraph}
\end{figure}

Figure \ref{fig:ConstraintGraph} is the
constraint graph of the scenario in Figure \ref{fig:threeFaust}. The \texttt{ntcc} model is
parametric on the constraint graph, which can be obtained from the
abstract semantics of the score, introduced in \cite{tdcr12}. High
precision relations are represented
as 
zero durations in the constraint graph because they are
controlled by Faust and not by Ntccrt, even if the duration in the
relations were given in seconds.

\begin{figure}[!h]
 \centerline{
 \includegraphics[width=\columnwidth]{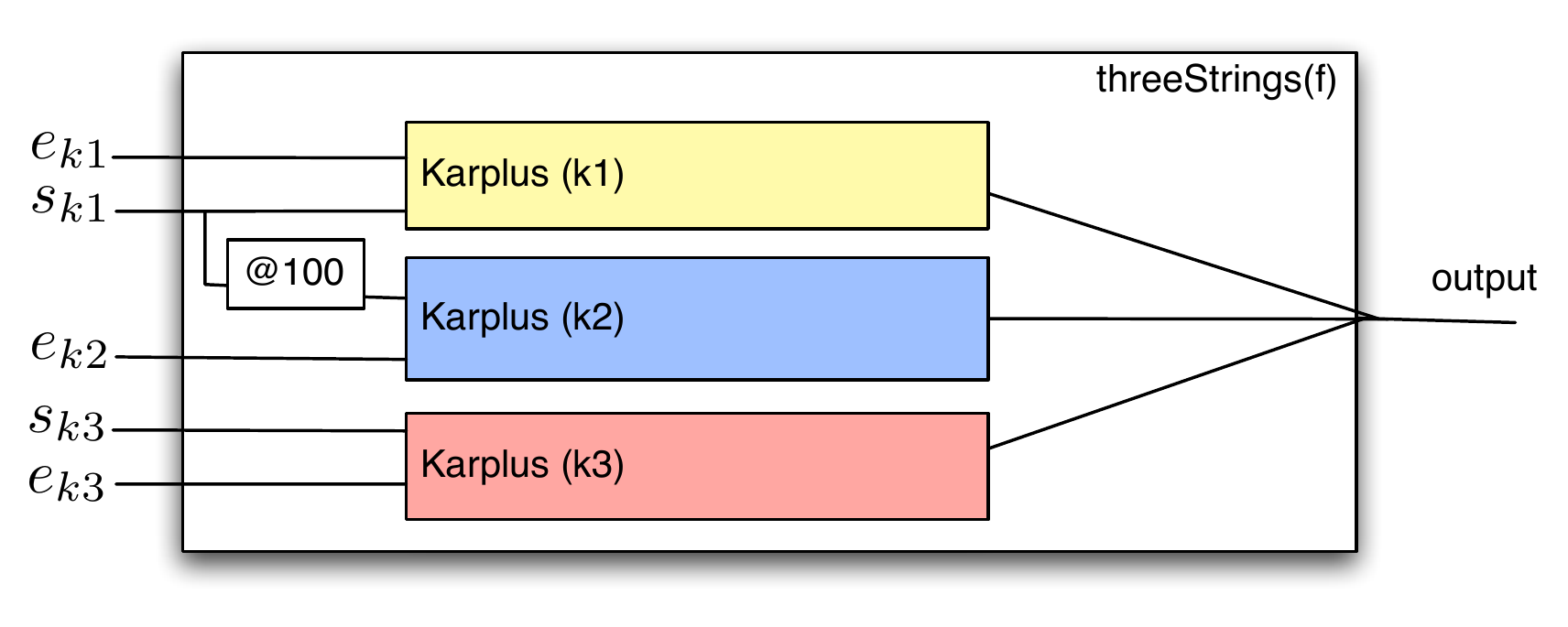}}
 \caption{Block diagram representing the Faust process in charge of
   signal processing and the micro controls of the sound processors of
   the scenario. Signal
   processor @100 adds a delay of 100 samples to the signal $s_{k1}$ (the start of the first string).}
 \label{fig:FaustDiagram}
\end{figure}

\begin{figure}[!h]
 \centerline{
 \includegraphics[width=6cm]{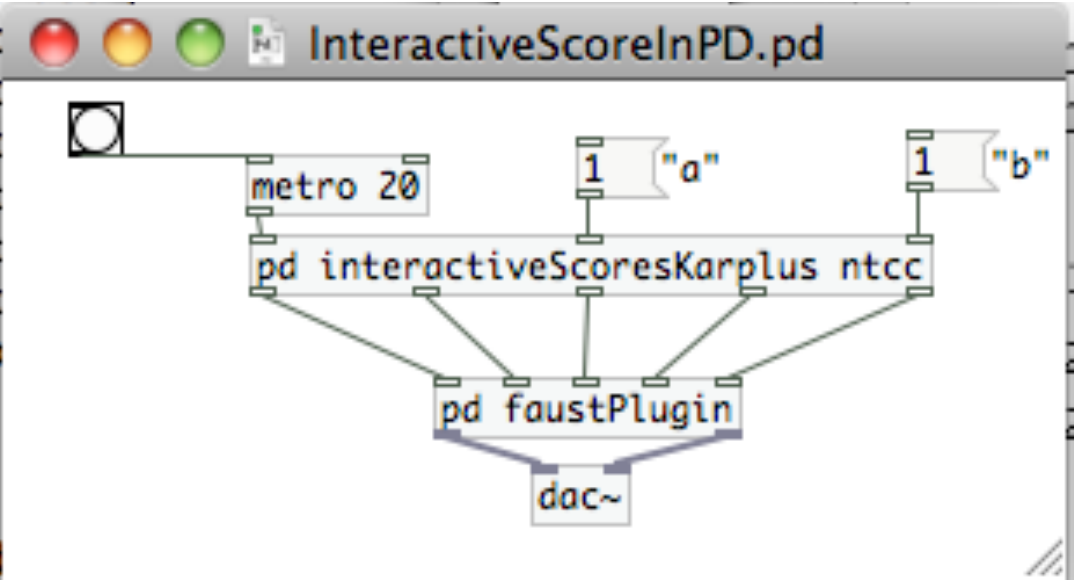}}
 \caption{Pure Data patch representing the scenario in Figure \ref{fig:threeFaust}. The Ntccrt plugin has
   only five outputs because the start of the second
   Karplus-Strong object ($k2$) is controlled
   directly from Faust. The internal clock of Ntccrt is controlled by a Pd
   \textit{metronome} object with a period of 20 ms.}
 \label{fig:ntccFaustPD}
\end{figure}

Figure \ref{fig:FaustDiagram} is the block diagram for the Faust
program in charge of sound processing. Block diagram semantics are
explained in detail in \cite{faust}. The inputs are controlled by the Ntccrt 
plugin. For simplicity, to avoid upsampling, control signals
(e.g., $e_{k1}$, $s_{k1}$ and $e_{k2}$) are replaced by Faust \textsc{gui} buttons
(Fig. \ref{fig:FaustButon}).  Interactive objects are represented by
   messages labeled by $1$: If a message arrives, the 
   interactive object must be launched. The audio output of each Karplus
block is added together into a single output.
Figure  \ref{fig:ntccFaustPD} is the Pure Data patch representing the
scenario. 

\section{Applications}
\label{applications}
We present some multimedia scenarios modeled in the
extended formalism of interactive scores.
 
\subsection{The Macro Structure of an Arpeggio Sequence }
\begin{figure}[!h]
 \centerline{
 \includegraphics[width=\columnwidth]{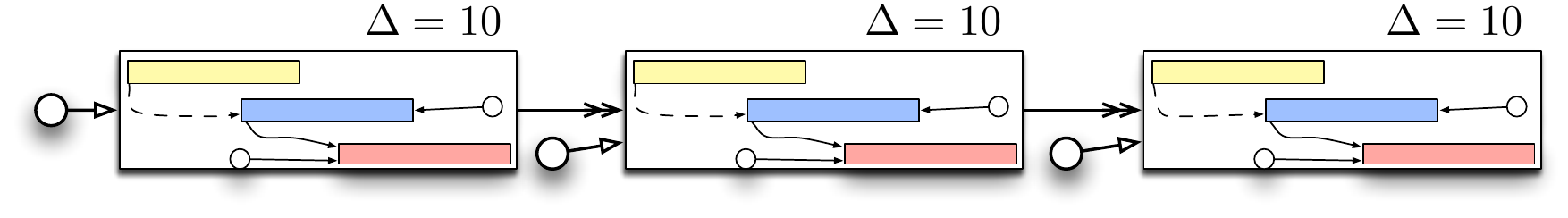}}
 \caption{Three repetitions of a temporal object containing an arpeggio of three
   strings (described in Figure \ref{fig:threeFaust}). The double-headed
   arrow represents an inequality ($\leq$) and a white-headed arrow
   represents an equality relation ($=$).}
 \label{fig:threeThreeFaust}
\end{figure}

In Figure
\ref{fig:threeThreeFaust}, we duplicate an arpeggio three times. The
macroform is respected: The duration of each arpeggio
is 10 seconds, 
but the start date and the 
durations of some notes can be controlled by the user
with the freedom described in Figure  \ref{fig:threeFaust}. This
problems shows how to solve the problem of having both time models
(the cue list and the fixed timeline models) temporally related. In
our framework, we can model the macroform of the arpeggio (e.g., the
duration of the 
notes and the global duration) and we can also model the microform
(e.g., the microdelays handled by Faust and the delays among the notes that can be controlled by user
interactions). 

\subsection{An Arpeggio without ``Clicks''}
There is a problem with the example in Figure \ref{fig:threeFaust}:
Interrupting abruptly the execution of the Karplus-Strong \textsc{dsp} 
causes perceptible ``clicks''. A solution to
this problem is to gradually decrease the volume (or increase the
attenuation parameter) 
 before stopping the \textsc{dsp}. The value of 0.5
 seconds 
is arbitrary, but it is fixed in the scenario, allowing us to know
precisely the macroform of the scenario (e.g., its total
duration). Therefore, instead of increasing the attenuation parameter
indefinitely, we represent the attenuation with a temporal object, thus
we can predict its duration and the global duration of the arpeggio.


\begin{figure}[!h]
 \centerline{
 \includegraphics[width=\columnwidth]{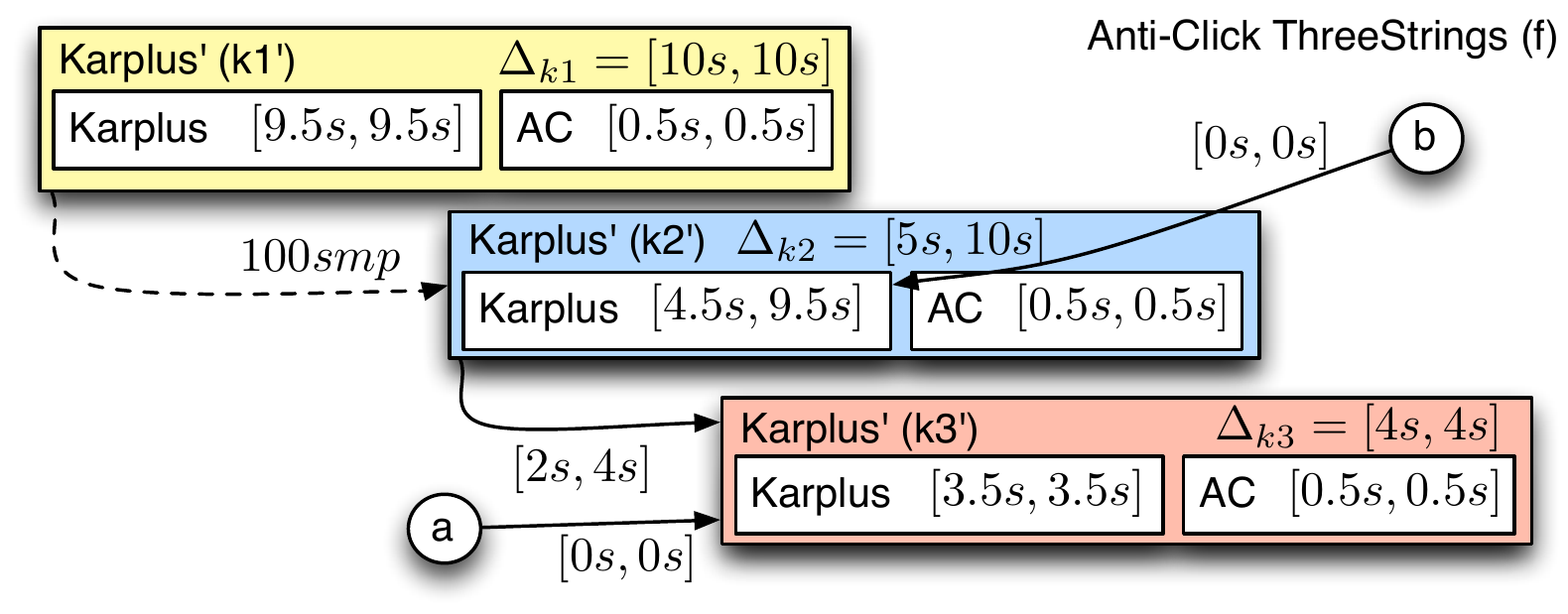}}
 \caption{A modification of the scenario, presented in Figure
   \ref{fig:threeFaust}, to remove ``clicks''. The Karplus objects
   simulate plucked-strings and the \textsc{ac} objects change the attenuation
   parameter of the strings gradually. The macroform of the scenario
   in Figure \ref{fig:threeFaust} is
   preserved intact.}
 \label{fig:antiClickScore}
\end{figure}

\subsection{Changing the Sound Source Perception}
\begin{figure}[!h]
 \centerline{
 \includegraphics[width=6cm]{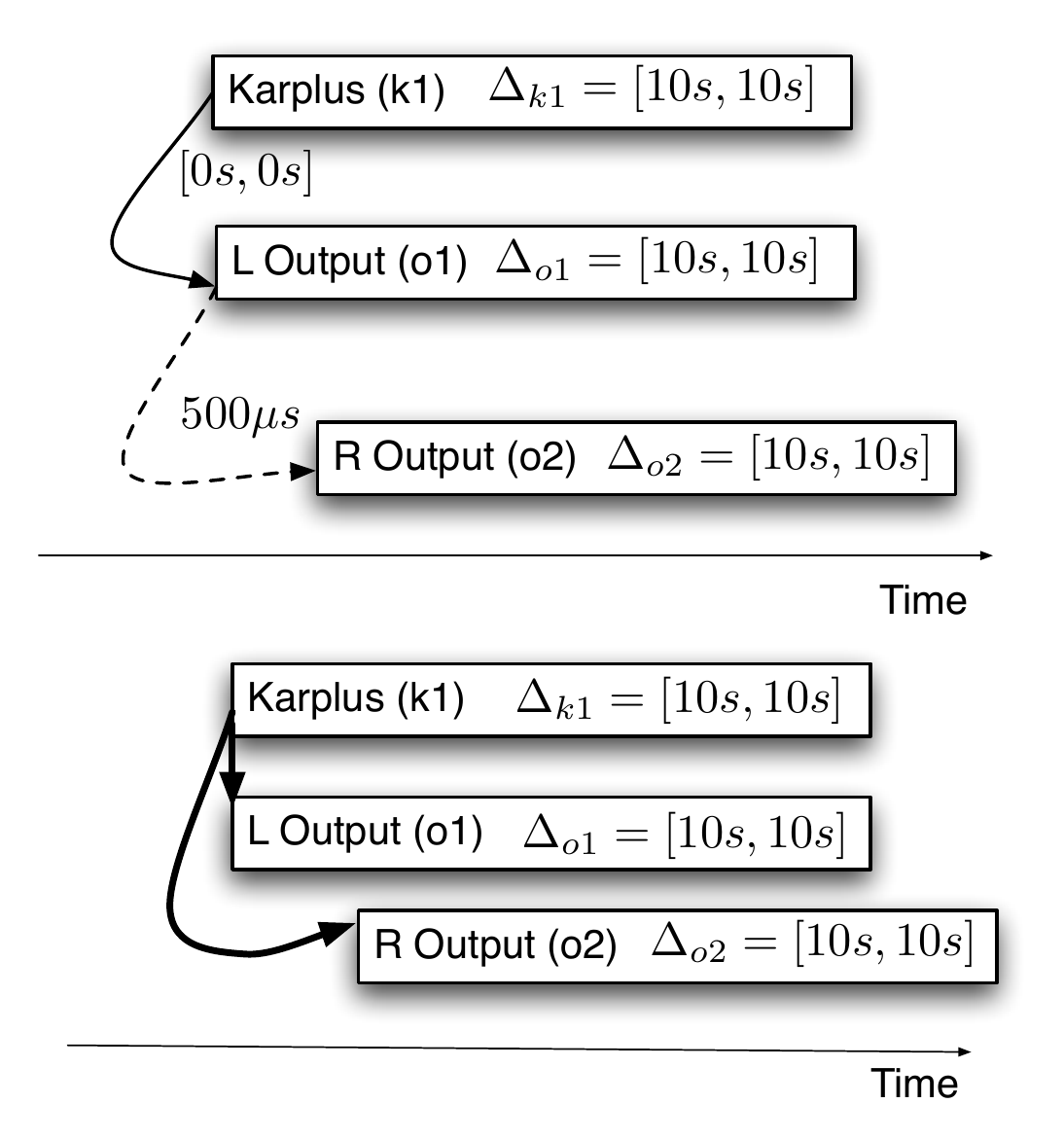}}
 \caption{A scenario with a micro interval. First output is the left channel
   and second output is the right channel. First view is
   \textit{temporal relations} and the second view is \textit{dataflow
   relations}. It is better to represent separately both views of the
score; otherwise, arrows will overlap.}
 \label{fig:MicroDelayScore}
\end{figure}

Small delays between the start of two temporal objects are usually
not perceptible; however, in some cases --such as the example in 
Figure \ref{fig:MicroDelayScore}--, a small delay of 500 $\mu
s$\footnote{This delay is equivalent to 22 samples at 44.1 kHz
  sampling rate.} between a sound
played on the left channel and the same sound played on the right channel
can change the way on which we perceive the sound source\footnote{http://buschmeier.org/bh/study/soundperception/}.

\section{Results}
\label{results}

We implemented the  arpeggio of
Figure \ref{fig:threeFaust}. We tested three implementations of the
Karplus-Strong in Pure Data (Pd): one from Colin Barry\footnote{www.loomer.co.uk}
that uses an instruction to define blocks of one sample (object
$block$\textasciitilde$\ 1$), one from Johannes
Kreidler\footnote{www.pd-tutorial.com} that uses one-sample delays
(object $z$\textasciitilde$\ 1$), and one from Albert Gr\"{a}f using a
Faust plugin generated
with Pd-Faust\footnote{http://docs.pure-lang.googlecode.com/hg/faust2pd.html}. The interactive objects are launched automatically (at the latest possible time).

For each test, we played each arpeggio four times with a \textsc{cpu}
load of 3\% and four times with a load of 85\%. We repeated each test
ten times. The tests were performed in a 3.06 GHz Intel Core i3 processor
on an iMac with a 
\textsc{ram} memory of 4 Gb 1333 MHz DDR3, under Mac
\textsc{os} 10.6.8, using Pure Data extended 0.42 and Faust 0.9. To increase the \textsc{cpu} load, we ran several video
processing operations from the \textit{graphics environment for
  multimedia} (\textsc{gem}) plugin for Pd. The \textsc{cpu} load
values are approximatively and they were obtained using Mac \textsc{os
  x}'s 
\textit{activity monitor}.

We calculated the average relative jitter of the micro- temporal
structure of the scenario: the average time difference between the expected starting time of each
string, with respect to the first string of the arpeggio, and the time
obtained during execution. The average
relative jitter using Faust is 500 $\mu$s with both a \textsc{cpu} load of 3\%
and 85\%; on the contrary, the implementation from Colin Barry has a
jitter of 7991 ms with a \textsc{cpu} load of 85\% and the
implementation from Johannes Kreidler has a jitter of 9231 ms with a
\textsc{cpu} load of 85\%. These values are very big and make the
listening of the arpeggio incomprehensible. The average
relative jitter was calculated using Matlab.

The Pd implementations of Karplus-Strong have also a limitation for high
frequencies: They work well until 2000 Hz and Faust works well until 3000 Hz.
Although this last result is the authors perception, we believe that the upper fundamental frequency limit may be due to
the ``chunk-sized'' buffer delay in the feedback loop in Pd. 

Another advantage 
of Faust is that the 
control signals in Faust can be delayed at sample level, whereas it is not 
possible to add sample delays to messages in Pd. In Pd, we need to delay the audio
output instead of the control signals to produce such result.
Finally, using Faust, sound processors could be automatically
parallelized, improving its performance in many cases \cite{lof10}.


\section{Conclusions}
\label{conclusions}
In this paper we extend the formalism of interactive scores with 
sound processing and micro controls for sound processors. We present an encoding of the
scenario into a \texttt{ntcc} model --executed using the real-time
capable interpreter \textit{Ntccrt}-- and a Faust program. Both programs interact 
during the performance of the scenario. We show how some interesting
applications can be easily modeled in the formalism and how they can be
executed in Pure Data (Pd).

 Using Faust and Ntccrt, we achieved an efficient and real-time capable performance
of a scenario --even under high \textsc{cpu}-load. Nonetheless, our final goal is to
integrate Ntccrt and Faust in a standalone program. 
 We argue that the solution we propose solves three of the problems we
posed in the introduction.

First, time models are related temporally, for instance, we can specify that an object is executed strictly in the third second of execution, and we can
can also express that another object is executed between two and five seconds after the end of the previous object.
Although in the execution the micro controls are managed by
Faust and the macro controls by \texttt{ntcc}, it is also possible to express, for instance, that an object starts 500 microseconds after another, and it will end
one second before another object.

Second, hierarchy is available in our model and it allows to constrain the execution times of the objects contained in another object.

Third, the system is appropriate, even under high \textsc{cpu}-load, to interact with a human in real-time, as shown in the quantitative results.

Unfortunately, different time scales are available in our 
tool, but they are temporally unrelated, as in many tools; for instance, is not
possible to relate the frequency of the clock that controls ntcc discrete time units to
the signal processing sampling rate.

Note that the score in Figure \ref{fig:threeFaust} is difficult to model in
 the existing tools presented in the introduction. Qlab and Live do not
allow to model delays of 100 samples. Max and Csound allow to express
delays of 100 samples, but it is very hard to synchronize processes
whose durations are 
integer intervals such as $duration \in [5,10]$.

The solution to these problems is relevant for the multimedia interaction domain
because, in addition to sound processing, the computer may execute at
the same time complex video and image operations. For that reason,
we did the evaluation of our system under  high \textsc{cpu}-load,
obtained by executing several video processing operations concurrently.

\subsection{Future Work}

We believe that any Faust program could be translated into \texttt{ntcc} based
on the results obtained by Rueda \textit{et al.} in \cite{audiontcc}.
 Rueda \textit{et al.} 
translated the Karplus-Strong Faust program into \texttt{ntcc}. Although it
is clear that the execution of a Ntccrt simulation cannot be done at sound
processing sampling frequency, such translation could be used to verify properties of
correctness of a scenario where \texttt{ntcc} and Faust interact
(e.g., playability) as proposed in \cite{iclp2010, audiontcc}.

We also propose to extend the implementation to handle audio
files efficiently. 
\textit{Libaudiostream}\footnote{http://libaudiostream.sourceforge.net/} is an 
audio library, developed at the french research institute
\textit{Grame}\footnote{http://www.grame.fr/}, to manipulate audio resources
through the concept of streams using  
Faust programs. 

Including Libaudiostream in our framework, it will be possible
to design a scenario where a temporal object loads a sound file into
memory,
Faust filter it, and then, Faust plays the 
sound at the appropriate time. Precision is guaranteed because the time to load the file and process it
is foreknown in the scenario. Currently, we have to rely on
third-party
programs, such as Pd, to do handle audio files, and to communicate the
control signals from Ntccrt to Faust.


\section*{Acknowledgments}
Ommited during blind reviewing



\small{
\bibliographystyle{cys}
\bibliography{mybib}

\begin{thebibliography}{10}
\expandafter\ifx\csname urlstyle\endcsname\relax
  \providecommand{\doi}[1]{doi:\discretionary{}{}{}#1}\else
  \providecommand{\doi}{doi:\discretionary{}{}{}\begingroup
  \urlstyle{rm}\Url}\fi

\bibitem{allen83}
\textbf{Allen, J.~F.} (\textbf{1983}).
\newblock Maintaining knowledge about temporal intervals.
\newblock \emph{Communication of ACM}, 26.

\bibitem{aad07}
\textbf{Allombert, A., Assayag, G., \& Desainte-Catherine, M.} (\textbf{2007}).
\newblock A system of interactive scores based on petri nets.
\newblock In \emph{Proc. of SMC '07}. Athens, Greece.

\bibitem{AADR06}
\textbf{Allombert, A., Assayag, G., Desainte-Catherine, M., \& Rueda, C.}
  (\textbf{2006}).
\newblock Concurrent constraint models for interactive scores.
\newblock In \emph{Proc. of SMC '06}. Marseille, France.

\bibitem{aadc08}
\textbf{Allombert, A., Assayag, G.~A., \& Desainte-Catherine, M.}
  (\textbf{2008}).
\newblock Iscore: a system for writing interaction.
\newblock In \emph{Proc. of DIMEA '08}. ACM, New York, NY, USA, 360--367.

\bibitem{virage}
\textbf{Allombert, A., Baltazar, P., Marczak, R., Desainte-Catherine, M., \&
  Garnier, L.} (\textbf{2010}).
\newblock Designing an interactive intermedia sequencer from users requirements
  and theoretical background.
\newblock In \emph{Proc. of ICMC 2010}.

\bibitem{is-chapter}
\textbf{Allombert, A., Desainte-Catherine, M., \& Toro, M.} (\textbf{2011}).
\newblock Modeling temporal constrains for a system of interactive score.
\newblock In \textbf{Assayag, G. \& Truchet, C.}, editors, \emph{Constraint
  Programming in Music}, chapter~1. Wiley, 1--23.

\bibitem{ArandaAOPRTV09}
\textbf{Aranda, J., Assayag, G., Olarte, C., P{\'{e}}rez, J.~A., Rueda, C.,
  Toro, M., \& Valencia, F.~D.} (\textbf{2009}).
\newblock An overview of {FORCES:} an {INRIA} project on declarative formalisms
  for emergent systems.
\newblock In \textbf{Hill, P.~M. \& Warren, D.~S.}, editors, \emph{Logic
  Programming, 25th International Conference, {ICLP} 2009, Pasadena, CA, USA,
  July 14-17, 2009. Proceedings}, volume 5649 of \emph{Lecture Notes in
  Computer Science}. Springer.
\newblock ISBN 978-3-642-02845-8, 509--513.

\bibitem{gennary98}
\textbf{Gennari, R.} (\textbf{1998}).
\newblock Temporal resoning and constraint programming - a survey.
\newblock \emph{CWI Quaterly}, 11, 3--163.

\bibitem{AG07}
\textbf{Gr\"{a}f, A.} (\textbf{2007}).
\newblock Interfacing pure data with faust.
\newblock In \textbf{LAC}, editor, \emph{5th International Linux Audio
  Conference (LAC2007)}.

\bibitem{jo11}
\textbf{Jouvelot, P. \& Orlarey, Y.} (\textbf{2011}).
\newblock Dependent vector types for data structuring in multirate faust.
\newblock \emph{Computer Languages, Systems and Structures}.

\bibitem{ntcc}
\textbf{Nielsen, M., Palamidessi, C., \& Valencia, F.} (\textbf{2002}).
\newblock Temporal concurrent constraint programming: Denotation, logic and
  applications.
\newblock \emph{Nordic Journal of Computing}, 1(9), 145--188.

\bibitem{cc-chapter}
\textbf{Olarte, C., Rueda, C., Sarria, G., Toro, M., \& Valencia, F.}
  (\textbf{2011}).
\newblock Concurrent constraints models of music interaction.
\newblock In \textbf{Assayag, G. \& Truchet, C.}, editors, \emph{Constraint
  Programming in Music}, chapter~6. Wiley, Hoboken, NJ, USA., 133--153.

\bibitem{faust}
\textbf{Orlarey, Y., Fober, D., \& Letz, S.} (\textbf{2004}).
\newblock Syntactical and semantical aspects of faust.
\newblock \emph{Soft Comput.}, 8(9), 623--632.
\newblock ISSN 1432-7643.
\newblock \doi{http://dx.doi.org/10.1007/s00500-004-0388-1}.

\bibitem{lof10}
\textbf{Orlarey, Y., Fober, D., \& Letz, S.} (\textbf{2010}).
\newblock Work stealing scheduler for automatic parallelization in faust.
\newblock In \emph{Proc. of Linux Audio Conference}.

\bibitem{PT13}
\textbf{Philippou, A. \& Toro, M.} (\textbf{2013}).
\newblock {Process Ordering in a Process Calculus for Spatially-Explicit
  Ecological Models.}
\newblock In \emph{{Proceedings of MOKMASD'13}}, {LNCS 8368}. Springer,
  345--361.

\bibitem{PTA13}
\textbf{Philippou, A., Toro, M., \& Antonaki, M.} (\textbf{2013}).
\newblock {Simulation and Verification for a Process Calculus for
  Spatially-Explicit Ecological Models}.
\newblock \emph{Scientific Annals of Computer Science}, 23(1), 119--167.

\bibitem{max}
\textbf{Puckette, M., Apel, T., \& Zicarelli, D.} (\textbf{{1998}}).
\newblock {Real-time audio analysis tools for Pd and MSP}.
\newblock In \emph{Proc. of ICMC '98}. Ann Arbor, USA.

\bibitem{audiontcc}
\textbf{Rueda, C. \& Valencia, F.} (\textbf{2005}).
\newblock A temporal concurrent constraint calculus as an audio processing
  framework.
\newblock In \emph{SMC '05}.

\bibitem{gecode}
\textbf{Tack, G.} (\textbf{2009}).
\newblock \emph{Constraint Propagation - Models, Techniques, Implementation}.
\newblock Ph.D. thesis, Saarland University, Germany.

\bibitem{tororeport}
\textbf{Toro, M.} (\textbf{2008}).
\newblock Exploring the possibilities and limitations of concurrent programming
  for multimedia interaction and graphical representations to solve musical
  csp's.
\newblock Technical Report 2008-3, Ircam, Paris.{(FRANCE)}.

\bibitem{torobsc}
\textbf{Toro, M.} (\textbf{2009}).
\newblock \emph{{Probabilistic Extension to the Factor Oracle Model for Music
  Improvisation}}.
\newblock Master's thesis, Pontificia Universidad Javeriana Cali, Colombia.

\bibitem{Toro-Bermudez10}
\textbf{Toro, M.} (\textbf{2010}).
\newblock Structured interactive musical scores.
\newblock In \textbf{Hermenegildo, M.~V. \& Schaub, T.}, editors,
  \emph{Technical Communications of the 26th International Conference on Logic
  Programming, {ICLP} 2010, July 16-19, 2010, Edinburgh, Scotland, {UK}},
  volume~7 of \emph{LIPIcs}. Schloss Dagstuhl - Leibniz-Zentrum fuer
  Informatik.
\newblock ISBN 978-3-939897-17-0, 300--302.
\newblock \doi{10.4230/LIPIcs.ICLP.2010.300}.

\bibitem{iclp2010}
\textbf{Toro, M.} (\textbf{2010}).
\newblock Structured musical interactive scores (short).
\newblock In \emph{Proc. ICLP 2010}.

\bibitem{torophd}
\textbf{Toro, M.} (\textbf{2012}).
\newblock \emph{{Structured Interactive Scores: From a simple structural
  description of a multimedia scenario to a real-time capable implementation
  with formal semantics }}.
\newblock Ph.D. thesis, Univerist{\'e} de Bordeaux 1, France.

\bibitem{Toro15}
\textbf{Toro, M.} (\textbf{2015}).
\newblock Structured interactive music scores.
\newblock \emph{CoRR}, abs/1508.05559.

\bibitem{ntccrt}
\textbf{Toro, M., Ag{\'o}n, C., Assayag, G., \& Rueda, C.} (\textbf{2009}).
\newblock {Ntccrt: A concurrent constraint framework for real-time
  interaction}.
\newblock In \emph{{Proc. of ICMC '09}}. Montreal, Canada.

\bibitem{tdc10}
\textbf{Toro, M. \& Desainte-Catherine, M.} (\textbf{2010}).
\newblock Concurrent constraint conditional branching interactive scores.
\newblock In \emph{Proc. of SMC '10}. Barcelona, Spain.

\bibitem{tdcb10}
\textbf{Toro, M., Desainte-Catherine, M., \& Baltazar, P.} (\textbf{2010}).
\newblock A model for interactive scores with temporal constraints and
  conditional branching.
\newblock In \emph{Proc. of Journ{\'e}es d'Informatique Musical (JIM) '10}.

\bibitem{tdcc12}
\textbf{Toro, M., Desainte-Catherine, M., \& Castet, J.} (\textbf{2012}).
\newblock An extension of interactive scores for multimedia scenarios with
  temporal relations for micro and macro controls.
\newblock In \emph{Proc. of Sound and Music Computing (SMC) '12}. Copenhagen,
  Denmark.

\bibitem{tdcr12}
\textbf{Toro, M., Desainte-Catherine, M., \& Rueda, C.} (\textbf{2012}).
\newblock Formal semantics for interactive music scores: A framework to design,
  specify properties and execute interactive scenarios.
\newblock \emph{Journal of Mathematics and Music. To appear in next volume}.

\bibitem{tdcr14}
\textbf{Toro, M., Desainte-Catherine, M., \& Rueda, C.} (\textbf{2014}).
\newblock {Formal semantics for interactive music scores: a framework to
  design, specify properties and execute interactive scenarios}.
\newblock \emph{Journal of Mathematics and Music}, 8(1), 93--112.
\newblock \doi{10.1080/17459737.2013.870610}.

\bibitem{mean-field-techreport}
\textbf{Toro, M., Philippou, A., Arboleda, S., V\'{e}lez, C., \& Puerta, M.}
  (\textbf{2015}).
\newblock {Mean-field semantics for a Process Calculus for Spatially-Explicit
  Ecological Models}.
\newblock Technical report, Department of Informatics and Systems, Universidad
  Eafit.
\newblock Available at
  http://blogs.eafit.edu.co/giditic-software/2015/10/01/mean-field/.

\bibitem{TPSK14}
\textbf{Toro, M., Philippou, A., Kassara, C., \& Sfenthourakis, S.}
  (\textbf{2014}).
\newblock Synchronous parallel composition in a process calculus for ecological
  models.
\newblock In \textbf{Ciobanu, G. \& M{\'{e}}ry, D.}, editors, \emph{Proceedings
  of the 11th International Colloquium on Theoretical Aspects of Computing -
  {ICTAC} 2014, Bucharest, Romania, September 17-19}, volume 8687 of
  \emph{Lecture Notes in Computer Science}. Springer.
\newblock ISBN 978-3-319-10881-0, 424--441.

\bibitem{toro-report09}
\textbf{Toro-Berm\'{u}dez, M.} (\textbf{2009}).
\newblock Towards a correct and efficient implementation of simulation and
  verification tools for probabilistic ntcc.
\newblock Technical report, Pontificia Universidad Javeriana.

\bibitem{Vickery04}
\textbf{Vickery, L.} (\textbf{2004}).
\newblock Interactive control of higher order musical structures.
\newblock In \emph{Proc. of ACMC}. ACMA, Victoria University, New Zealand.

\end{thebibliography}
}

\normalsize

\begin{biography}[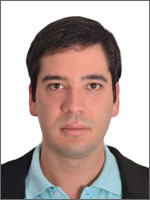]{Mauricio Toro} 
​is associate professor at the Computer Science Department at Universidad Eafit. He holds BsC, PhD and Postdoctorate degrees in Computer Science. His research field is simulation and verification of complex concurrent systems; in particular, he has worked on models of computer music interactive systems and ecological systems. He also holds an habilitation to supervise PhD students from the Doctorate of Engineering  Programme at EAFIT University. He has two research interests. ​​​First, He is interested in the simulation and verification of spatially-explicit individual-based ecological models for large populations. This was the topic of research he worked on during his Postdoctorate.
Second, He is interested in the simulation and verification of structured interactive music scores. This was the topic of research he worked on during his PhD.
\end{biography}

\begin{biography}[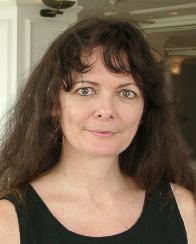]{Myriam Desainte-Catherine} 
is full professor at Université de Bordeaux and the Computer Science Research Laboratory of Bordeaux (LABRI). She has worked for more than 15 years on the formalism of interactive scores. She has supervised more than 10 PhD Students on different topics including sound and music computing. She is currently head of the music modeling team at LABRI and the Studio of Creation of Electroacoustic Music of Bordeaux  (SCRIME). 
\end{biography}

\begin{biography}[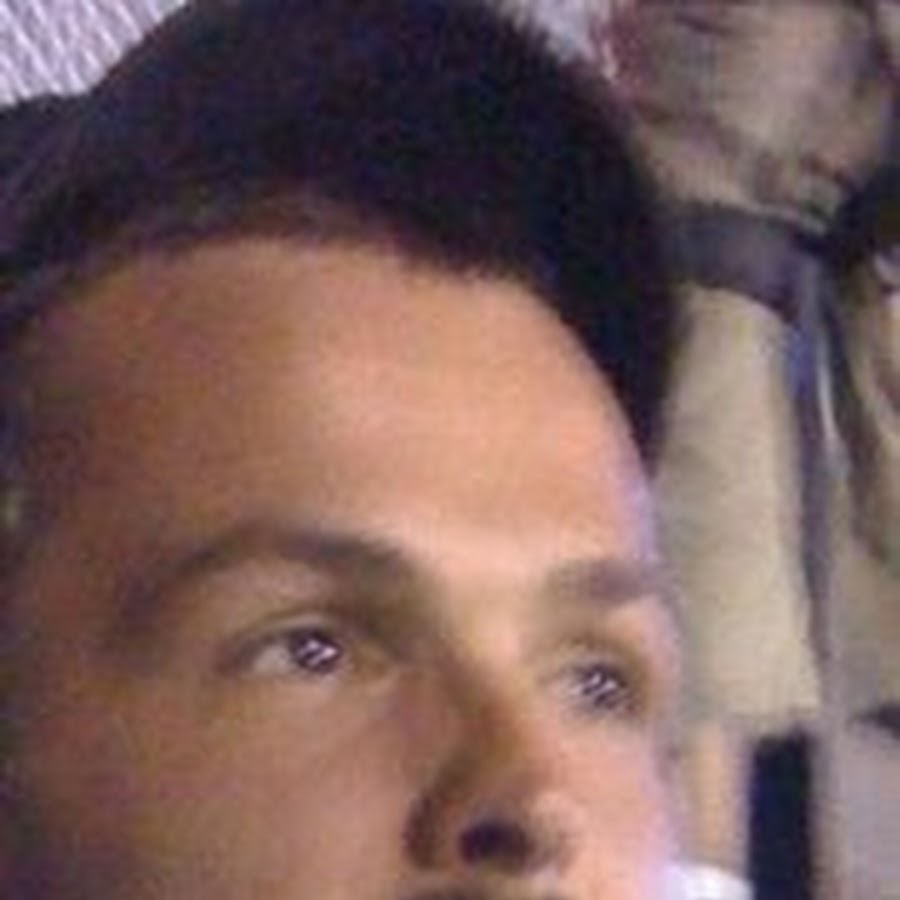]{Julien Castet} 
is researcher at Université de Bordeaux and the Computer Science Research Laboratory of Bordeaux (LABRI). He completed a PhD in the Music Modeling team at LABRI. His expertise is on signal processing and haptic instruments. 
\end{biography}

{\vskip 12pt}
\noindent
\footnotesize {Article received on 11/10/2015.}

\end{document}